\documentstyle [preprint,prl,aps] {revtex}
\begin{document}
\draft
\begin{titlepage}
\preprint{\vbox{ \hbox{IFP-472-UNC} \hbox{PRL-TH-93/13}
\hbox{hep-ph/9308348} \hbox{August 1993} }}
\title{ \large \bf Common Origin for the Solar
and\\Atmospheric Neutrino Deficits}
\author{Anjan S. Joshipura,}
\address{Theory Group, Physical Research Laboratory, Navrangpura,
Ahmedabad 380009, India}
\author{P.I. Krastev\footnote{Permanent address: Institute of Nuclear
Research and Nuclear Energy, Bulgarian Academy of Sciences,
BG--1784 Sofia, Bulgaria.}}
\address{Institute of Field Physics, Department of Physics and Astronomy,
The University of North Carolina at Chapel Hill, CB -- 3255, Phillips
Hall, Chapel Hill, NC 27599-3255}
\maketitle
\begin{abstract} Large mixing induced $\nu_{\mu}\leftrightarrow\nu_{e}$
transitions can explain the deficit in the flux of the low energy
atmospheric $\nu_{\mu}$ whereas the solar neutrino deficit can be
simultaneously explained through MSW
transitions $\nu_e\rightarrow\nu_{\tau}$.  A combined analysis of
both these effects is presented.
The large $\nu_{e}-\nu_{\mu}$ mixing affects
the MSW transition between $\nu_{e}$ and $\nu_{\tau}$ significantly.
As a consequence, a large region of parameters ruled out by experiment
in the two generation case is now allowed. The mass hierarchy as
well as the mixing pattern required arise naturally in a model for
neutrino masses proposed by Zee.
\end{abstract}
\end{titlepage}
\newpage
Two independent sets of experiments point towards the existence of neutrino
oscillations. The flux of the solar neutrinos is found to be depleted by
varied amounts compared to the expectations based on the standard solar
model in four different experiments namely, Homestake \cite{cl}, Kamioka
\cite{kam}, GALLEX \cite{gallex} and SAGE \cite{sage}. Similarly, the flux
ratio of the low energy ($\leq$ GeV) $\nu_{\mu}$ to $\nu_e$ of atmospheric
origin is seen to be smaller than the theoretically expected one in
experiments by the Kamioka II \cite{kam2} and IMB \cite{imb1} group.
Independent measurements by the Soudan\footnote{The latest, still
preliminary result from the SOUDAN--II detector seems to confirm the
results of Kamiokande and IMB, albeit with larger statistical errors.}
\cite{soudan} and the Frejus collaborations \cite{frejus} are not
inconsistent with the above but the statistical significance of these
results is much less than the corresponding one in cases
 of Kamioka and IMB results.
 Theoretically, both these deficits could originate from the same
common source, namely  neutrino oscillations. However the range of
the relevant parameters ((mass)$^2$ difference $\Delta$
 and mixing angle $\theta$ ) required by the solar
and atmospheric neutrino data are quite different.

The solar neutrino deficit can be explained through neutrino
oscillations occurring either in vacuum \cite{vac} or in the
presence of solar matter \cite{msw}. The former requires $\Delta
\approx 10^{-10}$ (eV)$^2$, sin$^22\theta \approx 0.75 - 1.0$
\cite{vac} while the latter becomes important if either
  $\Delta \approx (5 - 10) \times 10^{-6}$ (eV)$^2$, $\sin^2 2
\theta \approx (0.2 - 0.9)$ or $\Delta \sin^22\theta \approx (3 -
4) \times 10^{-8}$ (eV)$^2$ corresponding respectively to the large
angle adiabatic and non adiabatic Mikheyev Smirnov Wolfenstein
(MSW) solutions~\cite{pal}.  In contrast, the deficit in the
atmospheric muon neutrinos can be explained by means of neutrino
oscillations \cite{pak1} when parameters assume typical values
$\Delta \approx 10^{-3} - 10^{-2}$ (eV)$^2$ and $\sin^2 2 \theta
\approx 0.5$.

 It is quite clear that a simple two generation picture used in
deriving above restrictions is inadequate to simultaneously explain
the solar and the atmospheric neutrino deficits. A natural
possibility is to consider more realistic three generation picture.
When all three neutrino flavors are involved, the vacuum
oscillations can account \cite{pak2} for both the deficits if
$|\Delta_{ij}|\equiv |m_i^2-m_j^2|$ ($i,j$=1,2,3) $\simeq (0.5 -
1.2)\times 10^{-2} $ (eV)$^2$ and at least two of the mixing angles
are large. On the other hand if $|\Delta| \le 10^{-5} $ eV$^2$,
then the matter effects inside the Sun become important and they
should be taken into account. There are two independent
possibilities in this case.

\noindent ({\em i}) The oscillations between $\nu_e$ and
$\nu_{\mu}$ generate the solar deficit while the reduction in the
low energy atmospheric neutrino flux could be due to the
$\nu_{\mu}$ oscillating to $\nu_{\tau}$. This would require a large
$\theta_{23}$ and $|\Delta_{23}| \approx 10^{-2} - 10^{-3}$
(eV)$^2$. In this case, the high energy muon neutrinos originating
in the atmosphere also undergo oscillations as they travel through
the Earth. Such oscillations have not been found
\cite{imb2,others}. This negative evidence combined with the
positive evidence for oscillations in case of the low energy
neutrinos \cite{kam,imb1} strongly constrain the relevant
parameters and a very narrow range $2\times 10^{-3}\leq |\Delta
_{23}| \leq 0.4 $ (eV)$^2$, $0.4\leq \sin^2 2 \theta_{23}\leq 0.7 $
is consistent with all the available information\footnote{Allowed
regions of neutrino oscillation parameters solving the solar
neutrino problem in scenario~({\em i}) have been found in
ref.\cite{shi}.}\cite{akh}.\\

\noindent({\em ii}) The other possibility is to have a large mixing
between $\nu_e$ and $\nu_{\mu}$ and relatively large
$|\Delta_{21}|$ to generate the deficit in the atmospheric flux.
The MSW solution for the solar neutrino can be obtained in this
case if $\Delta_{31}$ falls in the appropriate range. The upcoming
high energy muons should oscillate in this case as well but these
oscillations will be affected by the presence of matter inside the
Earth.
 The values of parameters which are consistent with the
observations of the low and high energy atmospheric $\nu_{\mu}$
flux as well with the laboratory search for $\nu_e -\nu_{\mu}$
oscillations were determined to be \cite{akh}:  $4\times
10^{-3}\leq |\Delta _{12}| \leq 0.02 $ (eV)$^2$, $0.35\leq \sin^2 2
\theta_{12}\leq 0.7 $ .

In case ({\em i}), the solar and the atmospheric neutrino deficits
are explained through two independent sets of parameters and the
third generation does not play any more significant role than to
provide this set. In contrast, in case ({\em ii}), the presence of
a large mixing between the first two generations can influence the
values of $\Delta_{31}$ and $\theta_{13}$ allowed by the MSW
solution. The depletion of the solar neutrino is caused in this
case not only by the resonant conversion of the neutrinos inside
the Sun but also by their oscillations in vacuum arising due to
large mixing between the first two generations. Thus one expects
non-trivial departure compared to the case involving only two
generations. In this letter we study quantitatively the case~({\em
ii}).

The MSW mechanism in the case of three generations is more involved
because of the presence of the two independent (mass)$^2$
differences and two additional angles as compared with two
generations. However in a number of situations \cite{three}, one can
effectively reduce the problem to that of two generations. Case
({\em ii}) which requires vastly different values for $|\Delta_{31}|$
and $|\Delta_{21}|$ falls under this category.  One of the mass
eigenstates is expected to remain unaffected by matter and the
other two to undergo a resonant transition.
 This can be seen analytically by means of an approximation
\cite{kuo}. Let us assume that mixing between the second and the
third generation is very small and neglect it altogether. The
evolution of neutrino states inside the Sun is governed by a
(mass)$^2$ matrix which can be conveniently expressed in flavor
basis as follows:
\begin{equation}
 M_A^2= R_{12}(\theta_{12}) R_{13}(\theta_{13})
        \left( \begin{array}{ccc}
        0&0&0\\
        0&\Delta_{21}&0\\
        0&0&\Delta_{31}\\
        \end{array}\right)
       R_{13}^T(\theta_{13}) R_{12}^T(\theta_{12})+
\left(\begin{array}{ccc}
        A&0&0\\
        0&0&0\\
        0&0&0
        \end{array}\right),
\end{equation}
where we have subtracted
a term $m_1^2 I$ proportional to the identity matrix from the
actual (mass)$^2$ matrix. $R_{ij}$ is rotation in the $(ij)$ plane
described by the mixing angle $\theta_{ij}$.  $A=2\sqrt{2} G_F n E$
with $n$ and $E$ denoting the electron density inside the Sun and
the neutrino energy respectively. One could exactly diagonalize the
upper $2 \times 2$ block of this matrix by a rotation in the 12
plane by an angle $\theta_{12A}$.
\begin{equation} (M'_{A})^2\equiv
R_{12}^T(\theta_{12A})M_A^2 R_{12}(\theta_{12A});
\end{equation}
\begin{equation} \tan \theta_{12A}\equiv -\frac{2
(M_A^2)_{12}}{(M_A^2)_{11}-(M_A^2)_{22}}.
\end{equation}
We are
interested in the limit $|\Delta_{21}|\gg |\Delta_{31}|$ and $A$.
In this case, the $\theta_{12A}$ depends very mildly on $A$:
\begin{equation}
\theta_{12A}\approx\theta_{12}+\epsilon;\;\;\;\;\;\;\;
{\mbox{with}}\;\;\; \epsilon\equiv \sin 2\theta_{12}\frac{A}{2
\Delta_{21}}.
\end{equation}
In addition, the matrix
$(M^\prime_A)^2$ also assumes a simple form when non leading terms
of O($\epsilon$) are neglected: \begin{equation}
(M^\prime_A)^2\approx \left[
                      \begin{array}{ccc}
                      s_{13}^2\Delta_{31}+A c_{12}^2&0&s_{13}c_{13}
\Delta_{31}\\
                      0&\Delta_{21}& 0\\
                      s_{13}
c_{13}\Delta_{31}&0&\Delta_{31}c_{13}^2\\
                      \end{array}
                      \right] + O(\epsilon).
\end{equation}
where
$c_{ij}\equiv \cos\theta_{ij}$; $s_{ij}\equiv \sin\theta_{ij}$.  It
is seen from the above equation, that in the limit $\epsilon$ going
to zero, the second generation decouples and MSW resonance could
occur between the first and the third generation. The resonance
condition however differ in this case from the two generation case.
As follows from eq.(5), the resonance occurs if
\begin{equation}
 A \;\cos^2( \theta_{12})=\Delta_{31}\;\cos(2 \theta_{13}).
\end{equation}
For the range of $\Delta_{21}$ that is of interest
to us from the point of view of the atmospheric neutrino deficit,
$\epsilon\leq 10^{-2}$ and to a good approximation a resonance
occurs between only two of the neutrinos. The third mass eigenstate
remains more or less independent of $A$. This argument is confirmed
by the exact calculation of the variation of the effective neutrino
masses in matter with the solar density displayed in fig. 1 for
typical values of the parameters.

We shall work with eq.(5) with O($\epsilon$) terms neglected and
comment upon validity of such an assumption later on. The
probability $P_{\nu_e \nu_e}$ for the $\nu_e$ produced in the Sun
to remain $\nu_e$ at the detector follows immediately using eq.(5)
and the standard MSW picture \cite{pal}. Initially, the $\nu_e$
would be a mixture of three mass eigenstates determined by the form
of the mixing matrix $R$ at the point of production. One of the
mass eigenstates remains independent of $A$ while the other two
change and undergo resonant transition. Let $X$ denote the
probability of non-adiabatic transition between the two transforming
states.
The survival probability is then given by
\begin{equation} P_{\nu_e \nu_e}=\frac{1}{2}\left(
                                c_{12}^4(1 +(1-2 X) \cos(2
\theta_{13}) \cos(2 \theta_{13A_0}))+2 s_{12}^4 \right).
\end{equation}
As follows from eq.(4), the angle $\theta_{12A}$ is
mildly dependent upon $A$ and we have used its vacuum value in the
above equation. The $\theta_{13A_0}$ is the value of $\theta_{13A}$
at the production point. This can be determined from eq.(5)
\begin{equation}
\tan \theta_{13A_0}\equiv -\frac{\Delta_{31}\sin(2
\theta_{13})}
                               {\Delta_{31}\cos(2\theta_{13})-A_0
c_{12}^2}.
\end{equation}
The exact expression for $X$ depends upon
the assumed variation of the density in the vicinity of the
resonance layer. We shall use the following form which is valid
when this variation is linear but which provides a good
approximation to the actual situation in most cases \cite{pal}
\begin{equation}
X=\exp\left[ -\frac{\pi \Delta_{31} \sin^2
2\theta_{13}}
             {4\; E \cos(2\theta_{13}) \left|\frac{d}
             {d\;x}ln\;n\right|_R}\right].
\end{equation}
The above expression follows from a straightforward generalization
of the method used in case of purely two generation case
\cite{pal}. Unlike the resonance condition, eq.(6), $X$ turns out
to be independent of the large mixing $\theta_{12}$ between the
first two generations and retains its form given in purely two
generation case \cite{pal}.

  We have performed several numerical tests of the accuracy of the
analytical expressions for the electron--neutrino survival
probability given by eqs.(8). In particular, we have checked the
accuracy of this expression for several values of the parameters
$\Delta_{ij}$ and $\theta_{ij}$.  This has been done by comparing
the results of explicit numerical calculation of the survival
probability with the corresponding values given by eq.(8).  One
potential source of discrepancy between the values to be compared
is the nonequivalence of the two approaches. The analytical formula
presumes an averaging over all vacuum oscillation lengths. On the
other hand, the numerically obtained survival probability at the
surface of the Sun has still to be averaged over two oscillation
lengths $L_{12}$ and $L_{31}$. It is not clear a priori in which
order this averaging should be performed numerically.  We have
obtained good agreement between the analytical and numerical
results when we averaged first over the shorter oscillation length
$L_{12}$ and then over the longer one $L_{31}$. In this case
typically the relative error of the analytical formula does not
exceed several percent.

  Having checked numerically the accuracy of the analytical formula
we have proceeded with the analysis of the results from the solar
neutrino experiments in terms of three neutrino oscillations in
matter. We have used the latest available data from the four solar
neutrino experiments taking data at present. For the ratio of the
experimentally measured to the theoretically predicted event rate
in the chlorine experiment\cite{cl} we have used the value
\begin{equation}
R_{Cl} = 0.28 \pm 0.03.
\end{equation}
The relevant
ratio determined from the published results of the $Ga - Ge$
experiments SAGE\cite{sage} and GALLEX\cite{gallex} is
\begin{equation}
R_{\rm SAGE} = 0.44 \pm 0.21.
\end{equation}
\begin{equation}
R_{\rm GALLEX} = 0.66 \pm 0.12.
\end{equation}
We have used also the latest result from the $\nu_ee$ scattering
experiment conducted by the Kamiokande collaboration\cite{kam}
\begin{equation}
R_{Kamioka} = 0.49 \pm 0.08.
\end{equation}
In
our calculations we have used the solar model\cite{BP} which takes
into account the diffusion of hellium and is in excellent agreement
with the data from helioseizmology. The restrictions on the
parameters $\Delta_{31}$ and $\theta_{13}$ implied by the solar
neutrino experiments depend upon the value of $\theta_{12}$.
We have chosen values of $\theta_{12}$
allowed by the combined search for the $\nu_{\mu}$
oscillations in atmospheric as well as accelerator
neutrino experiments. These values, as determined in ref. \cite{akh},
fall in the range  $0.35\leq\sin^2 2 \theta_{12}\leq 0.7$.
They were determined for the case of two generations. But the
presence of the third generation is not expected to significantly
influence this determination as long as the $\nu_{\tau} -
\nu_{\mu}$ and $\nu_e - \nu_{\mu}$ mixings are small as assumed
here. We shall use two representative values of $\theta_{12}$ which
fall in the allowed band in the following.
 We also do not expect the results of the analysis of the solar
neutrino deficit with the effect of the Earth on the oscillations
of two neutrinos leading to unobserved seasonal and day--night
variations\cite{kamvar} to change considerably with the inclusion
of the third neutrino. The mixing angle $\theta_{13}$ relevant for
the solar neutrino oscillations is too small to affect the neutrino
oscillations in the Earth because of the much smaller dimensions of
the Earth as compared with that of the Sun. Also the oscillation
length corresponding to $\Delta_{31}$ is much larger then the
diameter of the Earth.

The neutrino survival probability for each set of neutrino
oscillation parameters has been computed by numerically finding the
resonant point and substituting in eq.9 the corresponding value of
the logarithmic derivative of the electron number density.  With
the neutrino survival probabilities so obtained we have calculated
the corresponding event rates in each detector and have compared
them with the ones predicted in the standard solar model.  The
allowed regions of parameters for which the neutrino conversion
hypothesis under consideration cannot be rejected at certain c.l.
are determined by minimizing the function:
\begin{equation}
\chi^2 = \sum_{i = Cl, Ga, H_2O}
\left({R_i^{exp} - R_{i}^{th}}\over{\sigma_i}\right)^2.
\end{equation}
They are shown in fig.2 for two different values of the large
mixing angle $\theta_{12}$.
Note that with diminishing the mixing angle $\theta_{12}$ the
allowed region of parameters $\Delta_{31}$ -- $\sin^22\theta_{13}$
converges to the one determined in the two--neutrino oscillation
scenario \cite{mswstat}.
As a less trivial result from the above calculation it
follows that a relatively large part of the adiabatic region of
parameters, i.e. the horizontal branch of the MSW triangle between
$\sin^22\theta_{13} = 0.07$ and $\sin^22\theta_{13} = 0.9$ is
allowed at the 95 \% c.l. for $\sin^22\theta_{13} = 0.65$.  The
mixed solution, i.e. the upper part of the hypotenuse of the MSW
triangle, is also allowed at the 90 \% c.l. for $\sin^22\theta_{12}
= 0.65$ and at the 95 \% c.l. for $\sin^22\theta_{12} = 0.45$. This
result is characteristic of the three--neutrino mixing scheme we
are considering here whereas in the ``standard'' two--neutrino
oscillation scenario the adiabatic and mixed solutions are ruled
out at the 95 \% c.l. from the combined analysis of the results of
the four solar neutrino experiments\cite{mswstat}.

These results change considerably when we include in our analysis
the recoil electron energy spectrum data obtained by the
Kamiokande--II collaboration. Instead of comparing only the mean
value of the suppression factor $R_{H_2O}$ with the predicted ones
for each set of neutrino oscillation parameters, we have deleted
the corresponding summand from eq.(13) and have added instead the
$\chi^2$ function for the 12 recoil electron energy bins as given
in\cite{kamsp}. We end up with the plot shown in fig.3. In this
case only a part of the so called ``mixed'' solution survives but
most of the adiabatic region is ruled out at the 95 \% c.l..
However, the allowed region of parameters is still different from
the corresponding one in the two--neutrino oscillation case.

Solar neutrino experiments that are in operation, as well as
proposed ones that are being developed will make it possible to
test the proposed three--neutrino oscillation solution of the solar
neutrino problem. For $\sin^22\theta_{12} = 0.65$ and
$\sin^22\theta_{13} = 5\times 10^{-4}$ and $\Delta_{31} = 8\times
10^{-5}$ eV$^2$ the signals in Ga-Ge detectors should be close to
82 SNU, in chlorine detectors about 2.52 SNU and the signal in
water Cherenkov-detectors should be about 0.33 of the predicted one
within the standard solar model. For $\sin^22\theta_{13} = 6\times
10^{-3}$ and $\Delta_{31} = 5\times 10^{-6}$ eV$^2$ the signals in
Ga-Ge and chlorine detectors should be correspondingly 53 SNU and
2.20 SNU and the suppression of the signal in $\nu_e e$ scattering
experiments should be 0.45.  Finally, for $\sin^22\theta_{13} =
0.7$ and $\Delta_{31} = 10^{-4}$ eV$^2$ the signals in these
detectors should be 55 SNU, 2.28 SNU with a suppression of 0.34 of
the signal in neutrino--electron scattering experiments.  The $pp$
neutrino signals alone should be suppressed by factors of 0.66,
0.63 and 0.44 correspondingly, whereas 0.862 MeV $^7Be$ neutrino
signals should be suppressed by factors 0.67, 0.47 and 0.44
respectively.
 Thus by comparing the signals in Borexino\cite{borexino},
SNO\cite{sno}, ICARUS\cite{icarus} and
Superkamiokande\cite{superkam} it will be possible to pinpoint the
allowed region of parameters.

 The proposed hellium detector of solar neutrinos
HELLAZ\cite{hellaz} is expected to measure the fluxes of both $pp$
and $^7Be$ neutrinos. Moreover, the spectrum of $pp$ neutrinos will
possibly be measured too.  As shown in\cite{bahcsp} this spectrum
does not depend on details of the solar model and its shape can be
predicted with a great precision.  Any significant deviation from
the standard shape will be an unequivocal evidence in favour of
neutrino conversion taking place either in the Sun, between the Sun
and the Earth and/or in the Earth.

The scenario we have considered requires a hierarchy
$|\Delta_{21}|\gg|\Delta_{31}|$. Such hierarchy does not occur in
seesaw models based on grand unified groups such as $SO(10)$ if the
right handed neutrino masses are assumed to be flavor independent.
However there exist other mechanisms for neutrino mass generation
which could lead to the required values for the $\Delta_{ij}$. A
concrete example is provided by the model of Zee \cite{zee} which
in fact comes very close to the present scenario. This model
contains two hierarchical $\Delta_{ij}$. The larger one
describes the $\nu_e-\nu_{\mu}$ oscillations while the
$\nu_e-\nu_{\tau}$ oscillations are of longer wavelength and are
controlled by the other (mass)$^2$ difference \cite{wolf}. The
model contains only left handed neutrinos whose majorana masses are
radiatively generated and are described by the following mass
matrix
\begin{equation} M_{\nu}=m_0 \left[
       \begin{array}{ccc}
        0&\sigma&\cos \alpha\\
        \sigma&0&\sin \alpha\\
        \cos \alpha&\sin \alpha&0
        \end{array}  \right],
\end{equation}
where $m_0,\sigma$ and $\alpha$ are parameters defined by Wolfenstein
\cite{wolf}. In the limit $\sigma$ going to zero, one has a
massless state and two with masses $\pm m_0$. We need to identify
the massless state with $"\nu_{\mu}"$. To first order in $\sigma$,
one has \cite{fn1} \begin{eqnarray} m_1&=&m_0(1+1/2 \sigma \sin 2
\alpha)\\ m_2&=&-\sigma \sin 2 \alpha\\ m_3&=&-m_0(1-1/2 \sigma
\sin 2 \alpha) \end{eqnarray} Hence, $\Delta_{21}\approx m_0^2$ ;
$\Delta_{31}\approx 2 m_0^2 \sigma \sin 2\alpha$. Thus our scenario
gets realized by choosing $m_0^2 \approx 10^{-3}$ (eV)$^2$ and $2
\sigma \sin 2\alpha \approx -10^{-2}$.

The mixing pattern predicted in the model also comes close to the
one required here. The mixing matrix is given in the limit $\sigma$
going to zero by:
\begin{equation} U M_{\nu} U^{T}=diag.
(m_1,m_2,m_3)
\end{equation}
\begin{equation} U^T=\left[
   \begin{array}{ccc}
   1/\sqrt 2 \cos \alpha&\sin \alpha&1/\sqrt 2 \cos \alpha\\
   1/\sqrt 2 \sin \alpha&-\cos \alpha&1/\sqrt 2 \sin \alpha\\
   1/\sqrt 2&0&-1/\sqrt 2
   \end{array}   \right].
\end{equation}
This coincides with the mixing matrix that we have used namely,
\begin{equation}
U^T=R_{12}(\theta_{12}) R_{13}(\theta_{13}),
\end{equation}
if one identifies $\theta_{12}=\pi-\alpha$ and
$\theta_{13}=\pi+\pi/4$. $\alpha$ and hence $\theta_{12}$ is not
fixed within the model but would be required to lie in the range
appropriate for solving the atmospheric neutrino problem. In
contrast, $\sin^2\theta_{13}$ is already fixed around\cite{fn2} its
maximal value.  This value is seen to be allowed by the
restrictions coming from the solar neutrino experiments as
displayed in figs.(2, 3). Thus, in addition to having the required
hierarchy in masses, the mixing angles in the model are also
consistent with the solar and atmospheric neutrino data.

{\bf Conclusions.} We have attempted to describe the observed
deficits in the solar and atmospheric neutrinos in terms of the
neutrino oscillations involving three generations. The deficits in
the flux of the low energy atmospheric $\nu_{\mu}$ is not as
clearly established \cite{perk} as the one observed in case of the
solar neutrinos. However if such a deficit gets firmly established
then the three generation scenario considered here would provide an
interesting mechanism for understanding both deficits in a coherent
manner. As discussed above, the conventionally employed two
generation MSW picture is still approximately applicable
in the present case but
the large mixing $\theta_{12}$ changes the allowed region of
parameters in a significant manner. Moreover, $\nu_{e}-\nu_{\mu}$
oscillations with relatively large (mass)$^2$ difference envisaged
here can be studied in laboratory as well.  Future experiments may
confirm or rule out the present scenario.

{\bf Acknowledgements.} A part of this work was done when A.S.J.
was visiting the Theory Division of CERN and the II. Inst. for
Theoretical Physics, Hamburg. He thanks G. Kramer and S. Pakvasa
and S. Rindani for discussions related to this work. The work of
P.K. has been partially supported by grant No. DE-FG05-85ER-40219
of the U.S. Department of Energy.  He thanks John Bahcall for a
number of very informative and fruitful discussions while visiting
the Institute for Advanced Studies, Y. Totsuka and Y. Takeuchi for
the detailed explanation of the procedure adopted by the Kamiokande
collaboration for the analysis of the recoil electron energy
spectrum and J.T. Liu for numerous discussions and invaluable help
concerning computer graphics.

\newpage
\centerline{\bf Figure Captions}
\noindent {\bf Fig.1} Masses of three flavors of
neutrinos as a function of $A$ expressed in units of 10$^{-5}$
eV$^2$ for the values of the vacuum parameters:
$m_1^2=0;\;m^2_2=500;\;m_3^2=1$ in units of $10^{-5}$ eV$^2$;
$\theta_{12}=40^0;\;\theta_{13}=5^0$ and $\theta_{23}=0.$
\vskip 0.5cm
\noindent {\bf Fig.2} Allowed region of parameters
$\sin^2(2\theta_{13})$ and $\Delta m_{31}$ at the 90 \% c.l. (solid
line) and 95~\% c.l. (dashed line) from the analysis of the results
of the four solar neutrino experiments.  The values of the mixing
angle $\theta_{12}$ have been chosen from the region allowed by the
analysis of the deficit of atmospheric $\nu_{\mu}$ in terms of
oscillations of two neutrino flavours in the Earth~\cite{kam2},
$\sin^2(2\theta_{12}) = 0.65$
 (fig.2a) and $\sin^2(2\theta_{12}) = 0.45$ (fig. 2b).

\vskip 0.5cm
\noindent {\bf Fig.3} The same as in fig.2 with the recoil electron
energy spectrum as measured by the Kamiokande-II collaboration
taken into account.


\newpage
\begin{thebibliography}{99}
\bibitem{cl} K. Lande {\it et al} in Proc. XXVth Int. Conf. on High
Energy Physics, Singapore, ed. K.K. Phua and Y. Yamaguchi, World
Scientific (Singapore 1991).
\bibitem{kam} Y. Totsuka, Talk given
at the ``Neutrino `92'' Int. Conf., Granada, June 7 -- 12, 1992;
Y. Suzuki, ICRR--Report--92--15.
 K.S. Hirata {\it et al} Phys. Rev. Lett. {\bf 66} (1991) 9.
\bibitem{sage} A. I. Abazov {\it et al} Phys. Rev. Lett. {\bf 67}
(1991) 3332;  V. Gavrin, talk at the Int. Conf. on High Energy
Physics, Dallas, 1992.
\bibitem{gallex} P. Anselman {\it et al,}
GALLEX collaboration report No.  GX 27a -- 1993 (to be published in
Phys. Lett. B).
\bibitem{kam2} K. S. Hirata {\it et al} Phys.
Lett. {\bf B 280} (1992) 146.
\bibitem{imb1} R. Becker-Szendy {\it
et al} Phys. Rev. {\bf D46} (1992) 3720; D. Casper {\it et al}
Phys. Rev. Lett. {\bf 66} (1992) 2561.
\bibitem{soudan} D. M.
Roback, Measurement of the atmospheric neutrino flavor ratio with
Soudan 2, Ph D Thesis , Univ. of Minnesota (1992).
\bibitem{frejus}
Ch. Berger {\it et al} Phys. Lett. {\bf B 245} (1990) 305;  {\bf B
227} (1989) 489.
\bibitem{vac} V. Barger {\it et al}, Phys. Rev.
{\bf D43} (1991) 1110; A. Acker, S. Pakvasa and J. Pantaleone,
Phys. Rev. {\bf D43} (1991) 1754; P. I. Krastev and S.T. Petcov,
Phys. Lett. {\bf B 299} (1993) 99.
\bibitem{msw} S.P. Mikheyev and
A. Yu. Smirnov, Yad. Fiz. {\bf 42} (1985) 1414; L. Wolfenstein,
Phys. Rev. {\bf D17} (1978) 2369.
\bibitem{pal} For a review see
P.B. Pal, Int. Journ of Mod. Phys. {\bf A7} (1992) 5387.
\bibitem{pak1} J. G. Learned, S. Pakvasa and T.J. Weiler, Phys.
Lett. {\bf B 207} (1988) 79.
\bibitem{pak2} A. Acker {\it et al},
Univ. of Hawaii Report, UH-511-746-92 (1992).
\bibitem{imb2} R.Becker Szendy {\it et al}, Phys. Rev. Lett.,
{\bf 69} (1992) 1010.
\bibitem{others} M. M. Boliev {\it et al}, in Proc. Third Int.
Workshop on Neutrino Telescop, ed. Milla Baldo Ceolin, (1991) p.
235; Y. Oyama {\it et al}, Phys. Rev. {\bf D39} (1989) 1481.
\bibitem{shi} X. Shi, D.N. Schramm and J.N. Bahcall, Phys. Rev.
Lett. {\bf 69} (1992) 717.
\bibitem{akh} E. Akhmedov, P. Lipari and
M. Lusignoli. Phys. Lett. {\bf B 300} (1993) 128.
\bibitem{three} For a review and references on MSW mechanism with
three generations, see T. K. Kuo and J. Pantaleone, Rev. of Mod. Phys.
{\bf 61} (1989) 937.
\bibitem{kuo} T. K. Kuo and J. Pantaleone,
Phys. Rev. Lett. {\bf 57} (1986) 1805.
\bibitem{BP} J. Bahcall and
M. Pinsonneault, Rev. Mod. Phys., {\bf 64} (1992) 885.
\bibitem{kamvar} K.S. Hirata et al., Phys. Rev. Lett. {\bf 66}
(1991) 9.
\bibitem{mswstat} P. Krastev and S. T. Petcov, Phys.
Lett. {\bf B 299} (1993) 99;  L. Krauss, E. Gates and M. White,
Phys. Lett. {\bf B 299} (1993) 94;  S. A. Bludman, N. Hata, D. C.
Kennedy and P. G. Langacker, Phys. Rev. {\bf D47} (1993) 2220;
see also A. Yu. Smirnov, Talk given at the International Symposium on
Neutrino Astrophysics, Takayama/Kamioka, 19 -- 22 October, 1992,
preprint ICTP/92/429.
\bibitem{kamsp} K.S. Hirata et al., Phys. Rev. Lett. {\bf 65}
(1990) 1301;  see also ``A solution to the Solar Neutrino Problem
by Matter--Enhanced Neutrino Oscillations'', Y. Fukuda, thesis,
Osaka University, 1992.
\bibitem{borexino} C. Arpesella et al.,
BOREXINO proposal, eds. G. Bellini, R. Raghavan et al. (Univ. of
Milan, Milan, 1992) vols. 1 and 2.
\bibitem{sno} G. Ewan et al.,
Sudbury Neutrino Observatory Proposal, SNO--87-12, 1987.
\bibitem{icarus} J.N. Bahcall, M. Baldo--Ceolin, D. Cline and C.
Rubbia, Phys. Lett. 178B (1986) 324; ICARUS I: An optimized,
real--time detector of solar neutrinos, by the ICARUS
collaboration, Frascati report LNF--89/005(R), 1989.
\bibitem{superkam} T. Kajita, Physics with the Super--Kamiokande
Detector, Tokyo preprint ICRR 185--89--2, 1989, Y. Totsuka, Tokyo
preprint ICRR 227--90--20, 1990.

\bibitem{hellaz} J.S\'eguinot, T.
Ypsilantis and A. Zichichi, Coll\'ege de France preprint LPC 92-31.
\bibitem{bahcsp} J.N. Bahcall, Phys. Rev. {\bf D44} 1992 1644, see
also V.A. Kuzmin and G.T. Zatsepin, Proc. of the 9th Int. Cosmic
Ray Conference, v.2, p.1023.
\bibitem{zee} A. Zee, Phys. Lett.
{\bf B 93} (1980) 389.
\bibitem{wolf} L. Wolfenstein, Nucl. Phys.
{\bf B175} (1980) 93.
\bibitem{fn1} The minis sign in one or two of
the masses can be removed by redefinition of the neutrino fields.
\bibitem{fn2} Inclusion of O($\sigma$) terms neglected here changes
the predicted mixing angles somewhat but in view of the small value
of $\sigma$ ($\leq O( 10^{-3}$) required here, this change is not
very large as can be seen from the explicit expressions given in
ref.\cite{wolf}.
\bibitem{perk} D. Perkins, Nucl. Physics {\bf
B399} (1993) 3.
\end{thebibliography}
\end{document}